\documentstyle[12pt]{article}

\newcommand{\al}{\alpha}

\newcommand{\df}{\stackrel{\rm def}{=}}

\newcommand{\msc}[1]{\mbox{\scriptsize #1}}
\newcommand{\dsp}{\displaystyle}

\newcommand{\br}{\mbox{{\bf R}}}
\newcommand{\bz}{\mbox{{\bf Z}}}

\newcommand{\bsz}{\msc{{\bf Z}}}
\newcommand{\bsr}{\msc{{\bf R}}}

\newcommand{\ket}[1]{{|#1\rangle}}

\newcommand{\Th}[2]{\Theta_{#1,#2}}
\newcommand{\th}{{\theta}}

\newcommand{\ch}[2]{\mbox{ch}^{#1}_{#2}}
\newcommand{\tch}[2]{\tilde{\mbox{ch}}^{#1}_{#2}}
\newcommand{\tr}{\mbox{Tr}}
\newcommand{\mod}{\mbox{mod}}

\newcommand {\eqn}[1]{(\ref{#1})}

\makeatletter
\@addtoreset{equation}{section}

\makeatother

\begin{document}

\vskip 7mm
\begin{titlepage}
 
 \renewcommand{\thefootnote}{\fnsymbol{footnote}}
 \font\csc=cmcsc10 scaled\magstep1
 {\baselineskip=14pt
 \rightline{
 \vbox{\hbox{hep-th/0002100}
       \hbox{UT-875}
       }}}

 \vfill
 \baselineskip=20pt
 \begin{center}
 \centerline{\Huge  Modular Invariance in Superstring} 
 \vskip 5mm 
 \centerline{\Huge on Calabi-Yau $n$-fold}
  \vskip 5mm 
 \centerline{\Huge with $A-D-E$ Singularity}

 \vskip 2.0 truecm

\noindent{\it \large Tohru Eguchi and Yuji Sugawara} \\
{\sf eguchi@hep-th.phys.s.u-tokyo.ac.jp~,~
sugawara@hep-th.phys.s.u-tokyo.ac.jp}
\bigskip

 \vskip .6 truecm
 {\baselineskip=15pt
 {\it Department of Physics,  Faculty of Science\\
  University of Tokyo\\
  Bunkyo-ku, Hongo 7-3-1, Tokyo 113-0033, Japan}
 }
 \vskip .4 truecm

 \end{center}

 \vfill
 \vskip 0.5 truecm

\begin{abstract}
\baselineskip 6.7mm

We study the type II superstring theory on the background
$\br^{d-1,1}\times X_n$, where $X_n$ is a  Calabi-Yau $n$-fold
($2n+d=10$) with an isolated singularity, 
by making use of the holographically dual description 
proposed by Giveon-Kutasov-Pelc \cite{GKP}. 
We compute the toroidal partition functions for each of the cases
$d=6,4,2$, and obtain manifestly modular invariant solutions 
classified by the standard $A-D-E$ series corresponding 
to the type of singularities on $X_n$. Partition functions of these modular 
invariants all vanish due to theta function identities and are
consistent with the presence of space-time supersymmetry.

\end{abstract}

\setcounter{footnote}{0}
\renewcommand{\thefootnote}{\arabic{footnote}}
\end{titlepage}

\newpage

\section{Introduction}

\hspace*{4.5mm}

String theory on singular backgrounds has been recently receiving
much attentions from various view points \cite{GKP}-\cite{hetero}. 
An important feature of a string propagating near
singularities is the  appearance of light 
solitons originating from  the branes wrapped around some 
vanishing cycles. This is a typical non-perturbative effect 
in string theory which is difficult to be worked out from
the world-sheet picture of perturbative string theory, 
even when a decoupling limit 
$g_s(\equiv e^{\phi(\infty)})\rightarrow  0$ is taken.
In fact, no matter how small  $g_s$ is,
the VEV of dilaton will blow up at the location of singularity.   
As was first pointed out in \cite{Aspinwall}, the 
{\em vanishing} world-sheet theta angle\/ ($\dsp \theta_{ws}\approx \int B=0$) 
is essential in the  appearance of such a non-perturbative effect. 
Several authors demonstrated \cite{Witten1,OV,Witten2,SW} 
that the conformal theory on string world-sheet
becomes singular in this situation.  On the other hand, 
the ordinary "smooth" conformal theory 
(orbifold CFT \cite{orbifold}, ${\cal N}=2$ 
Landau-Ginzburg model \cite{LG,LVW} etc.) 
corresponds to the backgrounds with a {\em non-vanishing} $\theta_{ws}$.
In the latter case the brane wrapped around a collapsed cycle
becomes a "fractional brane" \cite{fractional}  with a finite mass, 
and hence a perturbative approach to string 
theory is reliable,  at least if the string coupling $g_s$ is
sufficiently small and the mass of wrapped brane is large.

The first approach to such a singular CFT with a vanishing $\theta_{ws}$
was given in refs. \cite{GV,OV}
which were inspired by the theory 
of two dimensional black-hole. 
In these papers it is pointed out 
that such a singular conformal theory can be described 
by the Landau-Ginzburg (LG) model with a superpotential including a  
{\em  negative\/} power of some chiral superfield, 
and the subtlety in handling the negative power  
may be avoided by reformulating it as 
a Kazama-Suzuki model \cite{KazS} for the non-compact coset
$SL(2;\br)/U(1)$.   

More recently, a refinement of this approach was given 
in \cite{GKP,ABKS,GK,Pelc}, which is based on a holographic point 
of view analogous to the  
$AdS/CFT$ correspondence \cite{AdS}.
In these papers the sector of LG theory with 
a negative power superpotential is replaced 
by a suitable linear dilaton background 
(the ${\cal N}=2$ Liouville theory \cite{KutS2,KutS3})
describing the throat structure near the singularity, and 
it is pointed out that the  decoupled non-gravitational theory 
(on the space-time transverse to the singular Calabi-Yau $n$-fold)
has a dual description by a non-critical string theory 
including the dynamics of Liouville field \cite{KutS2,KutS3}. 
This duality is regarded as "holographic" in a manner
similar to $AdS/CFT$  in the sense that the throat variable 
(Liouville field) $\phi$ corresponds to an extra non-compact dimension,
and the decoupled theory is naturally defined in the weakly coupled
region $\phi \sim +\infty$ ("boundary").

This approach to the singular Calabi-Yau compactification
is interesting in the sense that the  non-critical 
superstring theory is playing a novel role. 
The main  purpose of this paper is to provide the basic consistency check
for these theories: the check of the modular
invariance of the toroidal partition function. 
In the case of $d=6$ (singular K3 surface) the result is straightforward.
The essential part of this case is already studied in \cite{OV},
and another approach from the standpoint of brane probe is given in \cite{DS}.
We have the standard $A-D-E$ classification of modular invariants
corresponding to the type of degeneration of 
K3 surface, which coincides exactly with 
the well-known modular invariants  of $SU(2)$ WZW model \cite{CIZ}.
This result is not surprising, since
the T-duality leads us to the theory of NS5-branes \cite{OV,GHM}  
and it is well-known \cite{CHS} that the world-sheet CFT 
of string propagation in the background of NS5-branes 
includes the $SU(2)$ WZW model.

The cases of $d=4$ and 2 are more difficult to analyse and 
are the central subjects of this paper. Although we do not have a simple 
world-sheet interpretation like the $d=6$ case,
we can construct and classify the modular invariants of these string theories. 
We will find out that the  conformal blocks in our models   
have modular transformation properties 
analogous to those of parafermion theories 
(coset CFT of $SU(2)/U(1)$) \cite{GQ} and 
we will again obtain the $A-D-E$ classification of modular invariants
corresponding to the type of singularities on $CY_n$.
It turns out that the partition functions of our modular invariants all 
vanish due to some theta function identities, which is  
consistent with the existence of space-time SUSY.

\bigskip

\section{Theory of Singular $CY_n$-Compactification as 
Non-critical Superstring}

\hspace*{4.5mm}

Let us consider  type II string theory on the background
$\br^{d-1,1}\times X_n$, where $X_n$ is a CY $n$-fold ($2n+d=10$)
with an isolated  singularity (locally) 
defined by $F(x^1,x^2,\ldots, x^{n+1})=0$.
As is pointed out in \cite{GVW,GKP}, in the decoupling limit
$g_s\,\rightarrow\, 0$ we have a non-gravitational,
but non-trivial  quantum theory on $\br^{d-1,1}$.
This fact  contrasts with the cases of a smooth $CY_n$, where
we expect a free theory in the $g_s\,\rightarrow\, 0$ limit. 
These $d$-dimensional quantum theories are expected to flow
to non-trivial conformal fixed points in the IR limit.
In the $d=6$ case, which is essentially the theory of NS5 branes,
they become the "little string theory" \cite{little,GK} 
including non-local excitations. The $d=4$ case corresponds
to the 4-dimensional ${\cal N}=2$ SCFT describing the Argyres-Douglas
points on the moduli space of ${\cal N}=2$ $SYM_4$ \cite{AD} 
(see also \cite{Pelc}). The $d=2$ case leads to the class of $AdS_3$
vacua with space-time ${\cal N}=2$ SUSY studied in \cite{GRBL}.
Space-time CFT is naturally identified with the boundary CFT in the context of 
$AdS_3/CFT_2$ correspondence.

The holographic duality proposed  in \cite{GKP} is represented  as follows;
$$
\begin{array}{l}
 \mbox{decoupling limit of superstring on } 
 \br^{d-1,1} \times X_n ~\Longleftrightarrow ~ \\
\hspace{4.5cm}
 \mbox{superstring on } \br^{d-1,1} \times (\br_{\phi}\times S^1) \times
LG(W=F),
\end{array}
$$
where $LG(W=F)$ stands for the ${\cal N}=2$ LG model with a 
superpotential $W=F$.
Moreover "
$\br_{\phi}$" indicates a linear dilaton background
with the background charge $Q (>0)$.
The throat sector $\br_{\phi}\times S^1$ is described by 
the ${\cal N}=2$ Liouville theory \cite{KutS2,KutS3}
whose field contents consist of bosonic variables 
$\phi$ (parametrizing $\br_{\phi}$), $Y$ (parametrizing  $S^1$)
and their fermionic partners $\Psi^+$, $\Psi^-$. The superconformal
currents are written as 
\begin{equation}
\left\{
\begin{array}{l}
\dsp  T =-\frac{1}{2}(\partial Y)^2 
-\frac{1}{2}(\partial \phi)^2 -\frac{Q}{2}\partial^2\phi
-\frac{1}{2}(\Psi^+\partial \Psi^- -\partial \Psi^+ \Psi^-) \\
\dsp  G^{\pm}=-\frac{1}{\sqrt{2}}\Psi^{\pm}(i\partial Y \pm \partial \phi )
\mp \frac{Q}{\sqrt{2}}\partial \Psi^{\pm} \\
\dsp J= \Psi^+\Psi^- -Qi\partial Y ,
\end{array}
\right.
\label{SCA-L}
\end{equation}
which generates the ${\cal N}=2$ superconformal algebra (SCA) with 
$\dsp \hat{c}(\equiv\frac{c}{3})=1+Q^2$.

Since we have a linear dilaton background, $\dsp \Phi(\phi) = -
\frac{Q}{2}\phi$, the theory is weakly coupled in 
the "near boundary region" $\phi\sim +\infty$.
On the other hand, in the opposite end $\phi \sim -\infty$ 
(near the singularity) the string coupling blows up,
and hence the perturbative approach does not make sense.
As is discussed in \cite{GKP}, one must add the "Liouville potential"
(or the "cosmological constant term")
to the world-sheet action of the Liouville sector,
\begin{equation}
\begin{array}{l}
 \dsp S_L ~\longrightarrow~ S_L +\delta S_+ + \delta S_- , \\
 \dsp \delta S_{\pm}\df \mu \int d^2z\, \Psi^{\pm}\bar{\Psi}^{\pm}
e^{-\frac{1}{Q}(\phi\mp i Y)} ,
\label{Liouville potential}
\end{array}
\end{equation} 
in order to prevent the string from propagating into
the dangerous region $\phi \sim -\infty$.
$\delta S_{\pm}$ are actually the screening charges in the sense
that they commute with all the generators of SCA \eqn{SCA-L}.
So, we shall carry out all the computations as a free conformal theory
on the world-sheet neglecting the existence 
of Liouville interaction \eqn{Liouville potential},
although we have to keep in our mind that we cannot set $\mu=0$\footnote
      {More rigorous setup may be the "double scaling limit"
        discussed in \cite{GK,Pelc}; 
  $$ g_s~ \rightarrow~0,~~~ \mu ~ \rightarrow~0 ~~~
  \mbox{with $\dsp \frac{\mu^{Q^2/2}}{g_s}$ fixed 
  to be a sufficiently large value},
  $$
  so that the theory is weakly coupled.}.

Throughout this paper we focus our attention to cases when $X_n$ has 
an isolated rational singularity.
In these cases the LG theory with $W=F$ is equivalent to 
the familiar ${\cal N}=2$ minimal models of the $A-D-E$ type corresponding 
to the classification  of rational singularities defined by $F=0$.
These models include the chiral primary fields which are in  one-to-one
correspondence with the exponents of the $A-D-E$ group, and 
have the central charge 
 $\dsp \hat{c}=\frac{N-2}{N}$, where $N$ is the dual Coxeter
number of the $A-D-E$ group. From now on we denote
these ${\cal N}=2$ minimal model as $M(G, N)$ $(G = A_m,\, D_m,\, E_m)$, or
more simply as $M_N$ when there is no problem of confusion. 
Hence, the model to be studied 
is the RNS superstring compactified on $\br^{d-1,1} 
\times (\br_{\phi}\times S^1)\times M_N$.  

The condition of critical dimension can be written as 
\begin{equation}
\frac{N-2}{N} + (1+Q^2) = n (\equiv \frac{10-d}{2}),
\label{criticality}
\end{equation}
and it is easy to evaluate $Q$
for each of the cases $d=6,4,2$
\begin{equation}
\begin{array}{ll}
  d=6; & \dsp Q=\sqrt{\frac{2}{N}},  \\
  d=4; & \dsp Q=\sqrt{\frac{N+2}{N}},\\
  d=2; & \dsp Q=\sqrt{\frac{2(N+1)}{N}}.
\end{array}
\end{equation} 

Notice that the criticality condition \eqn{criticality}
is equivalent to the Calabi-Yau condition 
for the non-compact $n$-fold (defined in a suitable 
weighted projective space)
\begin{equation}
\tilde{F}(x,z^1,z^2,\ldots,z^{n+1})
\equiv -\hat{\mu} x^{-\frac{2}{Q^2}} +F(z^1,z^2,\ldots,z^{n+1})=0.
\end{equation} 
In refs. \cite{GV,OV} the negative power term in the superpotential
$W\sim \hat{\mu}x^{-\frac{2}{Q^2}}$ is 
replaced by the Kazama-Suzuki model 
for $SL(2,\br)/U(1)$ with the level
$\dsp k'\equiv \frac{2}{Q^2}+2 $.
The equivalence between 
such a non-compact Kazama-Suzuki model
and the ${\cal N}=2$ Liouville theory \eqn{SCA-L} 
($\hat{\mu}$  corresponds to the cosmological 
constant $\mu$ in \eqn{Liouville potential}) was discussed in \cite{GK}
and it was pointed out that both theories are related 
by a kind of T-duality.  We argue for this equivalence from the
point of view of the free field realization in the Appendix B.

\bigskip


\section{Toroidal Partition Functions}

\hspace*{4.5mm}

We study the toroidal partition functions for the above non-critical 
superstring models.
The toroidal 
partition function  for RNS superstring
has the following general structure;
\begin{equation}
Z=\int \frac{d^2\tau}{\tau^2_2} \, 
Z_0(\tau,\bar{\tau})Z_{GSO}(\tau,\bar{\tau}),
\end{equation} 
where $\tau\equiv \tau_1+i\tau_2$ is the modulus of the torus 
($\dsp d^2\tau/\tau^2_2$ is the modular invariant measure).
$Z_{GSO}$ denotes the part of the partition function
which consists of those contributions on which the GSO projection acts
non-trivially. We write the remaining part as $Z_0$.

Obviously $Z_0$ includes only the contributions from the transverse
non-compact bosonic coordinates $\br^{d-2}\times \br_{\phi}$.
The Liouville sector $\br_{\phi}$ is slightly non-trivial because of 
the existence of the background charge. 
We should bear in our mind that only 
the normalizable states contribute to the partition 
function. The normalizable spectrum 
 (in the sense of the delta function 
normalization because the spectrum is continuous) 
in Liouville theory has the lower bound $\dsp h=\frac{Q^2}{8}$ 
\cite{KutS1,KutS3}. 
Since this lower bound is non-zero, we must be careful 
in the integration over the zero-mode momentum. 
The result, however, turns out to be the same as that of 
the standard non-compact free boson without background charge,
\begin{eqnarray}
&&Z_{L} (\tau,\bar{\tau}) 
= {1 \over |\prod_{n=1}(1-q^n)|^2}
\int_{-\frac{iQ}{2}-\infty}^{-\frac{iQ}{2}+\infty}dp 
\ \exp\left(-4\pi \tau_2\Big({1 \over 2}p^2+{i \over 2}pQ-{c_L \over 24}\Big)
\right), \nonumber \\
&&={1 \over |\prod_{n=1}(1-q^n)|^2}\int_{-\infty}^{+\infty}dp 
\ \exp\left(-4\pi \tau_2\Big({1 \over 2}p^2+{1 \over 8}Q^2-{c_L \over 24}\Big)
\right)
=\frac{1}{\tau_2^{1/2}|\eta(\tau)|^2}
\end{eqnarray}
where $c_L=1+3Q^2$. It is well-known \cite{KutS1} that the effective value 
of the Liouville central charge $c_{\msc{eff},L}$ is equal to 
\begin{equation}
c_{\msc{eff},L} \equiv c_L-24 \times \frac{Q^2}{8} =1 ,
\end{equation}  
irrespective of the value of $Q$ and the dependence on the 
background charge disappears from the net result.

In this way we obtain
\begin{equation}
Z_0(\tau,\bar{\tau})= \frac{1}{\tau_2^{(d-1)/2}|\eta(\tau)|^{2(d-1)}}.
\end{equation} 
This expression is manifestly modular invariant.

The part $Z_{GSO}$ is rather non-trivial. We need consider it
separately in the cases $d=6,4,2$.
We first discuss the simplest case $d=6$, and then proceed to 
the $d=2,4$ cases. 

\subsection{$d=6$ Case} 

Let ${\cal H}_{lm}^{(NS)}$ (${\cal H}_{lm}^{(R)}$) 
be the (left-moving) Hilbert space of the $NS$ (R) sector 
of CFT describing the minimal model $M_N$. 
The spectra of $U(1)_R$-charges and conformal weights  are
given by;
\begin{equation}
\begin{array}{lll}
 {\cal H}_{lm}^{(NS)}~(l+m\equiv 0 ~ (\mod~ 2)):~& 
 \dsp q=\frac{m}{N}+n ~(n\in \bz), & 
 \dsp h=\frac{l(l+2)-m^2}{4N}+n ~(n\in \frac{1}{2}\bz_{\geq 0}), 
 \\
{\cal H}_{lm}^{(R)}~(l+m\equiv 1 ~ (\mod~ 2)):~& 
 \dsp q=\frac{m}{N}-\frac{1}{2}+n ~(n\in \bz), 
& \dsp  h=\frac{l(l+2)-m^2}{4N}+\frac{1}{8}+n ~(n\in \bz_{\geq 0}). 
\end{array}
\end{equation} 

We also consider the (left-moving)
Fock space ${\cal H}_{p}$ of the bosonic coordinate $Y$
of $S^1$ constructed on the Fock vacuum  $\ket{p}$,
$\dsp \oint i\partial Y\ket{p}=p\ket{p}$. 
Values of the momenta $p$ 
are chosen so that they are compatible with the GSO projection condition.

The total ${\cal N}=2$ $U(1)_R$-charge is given by 
\begin{equation}
\begin{array}{lll}
 J_0^{(NS)}& = &\dsp F+(F_{M_N}+\frac{m}{N})-p Q, \\
 J_0^{(R)}& = &\dsp F+(F_{M_N}+\frac{m}{N}-\frac{1}{2})-p Q ,
\end{array}
\end{equation}
where $F$ denotes the fermion number of 
$\br^{d-2} \times (\br_{\phi}\times S^1)$ sector
and $F_{M_N}$ denotes the fermion number of the $M_N$ sector.

After these preparations conditions for the GSO projection
(the conditions for the mutual locality with the space-time 
SUSY charges) can now be written as 
\begin{itemize}
 \item $NS$-sector
\begin{equation}
F+F_{M_N}+\frac{m}{N}-p Q \in 2\bz +1,
\label{GSO-6-1}
\end{equation}
 \item $R$-sector
\begin{equation}
F+F_{M_N}+\frac{m}{N}-p Q \in 2\bz .
\label{GSO-6-2}
\end{equation}
\end{itemize}

Let us now compute the trace over the left-moving Hilbert space.
For instance, let us suppose $F+F_{M_N}=\mbox{even}$, 
and consider the $NS$-sector.
Then we have $\dsp p=\frac{1}{Q}\left(2n+1+\frac{m}{N}\right)$ 
$(n\in\bz)$, and the sum over momenta becomes,
\begin{equation}
\sum_n q^{{1 \over 2}p^2}=\sum_n q^{{N \over 4}(2n+1+{m \over N})^2}
=\sum_n q^{N(n+{m+N \over 2N})^2}=\Theta_{m+N,N}(\tau).
\end{equation}
When combined with factors coming from oscillator modes and the 
minimal model $M_N$, NS sector partition function becomes
$$
\frac{1}{2} \left\{\left(\frac{\th_3}{\eta}\right)^3\ch{(NS)}{lm}
+\left(\frac{\th_4}{\eta}\right)^3\tch{(NS)}{lm}\right\}\,
\frac{\Th{m+N}{N}}{\eta}.
$$ 
Here $\dsp \eta=q^{1/24}\prod_{n=1}^{\infty}(1-q^n)$ and  
$\dsp \th_3=\sum_n q^{\frac{n^2}{2}}$, 
$\dsp \th_4=\sum_n (-1)^n q^{\frac{n^2}{2}}$,
$\dsp \th_2=\sum_n q^{\frac{1}{2}\left(n-\frac{1}{2}\right)^2}$,
$\dsp \Th{m}{N}=\sum_n q^{N\left(n+\frac{m}{2N}\right)^2}$ 
$(q\equiv e^{2\pi i\tau})$ are the standard theta functions.
$\ch{(NS)}{lm}(\tau)$, $\tch{(NS)}{lm}(\tau)$ denote
the irreducible characters of ${\cal N}=2$
minimal model for $NS$-sector. 
(We summarize the definitions of these functions  in the appendix A.)
Similarly we can calculate the trace in other sectors, 
and obtain (we omit the factors of 
$\eta$-functions for simplicity),
\begin{eqnarray}
\sum_{l=0}^{N-2}\,  G_l&=& \frac{1}{2}\sum_{l=0}^{N-2} 
\sum_{m\in\bsz_{2N}}\left\{
\th_3^3\ch{(NS)}{lm}(\Th{m}{N}+\Th{m+N}{N})
-\th_4^3\tch{(NS)}{lm}(\Th{m}{N}-\Th{m+N}{N}) \right.  \nonumber\\
& & \hspace{1in} \left.
-\th_2^3\ch{(R)}{lm}(\Th{m}{N}+\Th{m+N}{N})
\right\}.
\label{Gl-6}
\end{eqnarray}
The above sum \eqn{Gl-6}, however, counts each state twice due to
the symmetry $G_l=G_{N-2-l}$.
To avoid this double counting, we may define 
\begin{equation}
F_l \df  \frac{1}{2}\sum_{m\in\bsz_{2N}}\Th{m}{N}
(\th_3^3\ch{(NS)}{lm}
-\th_4^3\tch{(NS)}{lm} -\th_2^3\ch{(R)}{lm} ). 
\label{Fl-6-1}
\end{equation}
and have
\begin{equation}
G_l = F_l+F_{N-2-l} .
\end{equation}
The desired partition sum then takes the form
\begin{equation}
Z_{GSO}(\tau,\bar{\tau})=\frac{1}{|\eta(\tau)|^8}
\sum_{l,\bar{l}=0}^{N-2}\, N_{l,\bar{l}}F_l(\tau)F_{\bar{l}}(\bar{\tau}).
\label{ZGSO-6}
\end{equation}  
Thanks to the branching relation \eqn{branching} we may rewrite 
$F_l$ in the following simple form \cite{OV};
\begin{equation}
F_{l}(\tau)=\frac{1}{2}
\left(\th_3^4-\th_4^4-\th_2^4\right)\chi^{(N-2)}_l(\tau)  ,
\label{Fl-6-2}
\end{equation}
where $\chi^{(k)}_l(\tau)$ denotes the $\widehat{SU}(2)_k$ character of the 
spin $l/2$ representation.
Hence the expression \eqn{ZGSO-6} is manifestly modular invariant 
when $N_{l,\bar{l}}$ is chosen to be one of the modular invariants
$L^{(N-2)}_{l,\bar{l}}$ of $\widehat{SU}(2)_{N-2}$ theory which fulfill the
conditions,
\begin{equation}
\begin{array}{l} \dsp
L^{(k)}_{l,\bar{l}}=0, ~~~\mbox{unless } 
       \frac{l(l+2)}{4(k+2)}-\frac{\bar{l}(\bar{l}+2)}{4(k+2)} \in \bz, ~~~
L^{(k)}_{k-l, k-\bar{l}}=L^{(k)}_{l,\bar{l}},\\
\dsp \sum_{l,\bar{l}}\, L^{(k)}_{l,\bar{l}}
S^{(k)}_{ll'}S^{(k)}_{\bar{l}\bar{l}'}=L^{(k)}_{l',\bar{l}'}, ~~~
\dsp S^{(k)}_{ll'}\df 
\sqrt{\frac{2}{k+2}}\sin \left(\pi \frac{(l+1)(l'+1)}{k+2}\right).
\end{array}
\label{modinvsu2}
\end{equation}
Furthermore, $F_l$ (\ref{Fl-6-2}) 
identically vanishes by virtue of the Jacobi's abstruse
identity: 
this is consistent with the existence
of space-time SUSY.\footnote
    {As discussed in \cite{KutS2,KutS3}, we only have the SUSY 
    along the "boundary" $\br^{d-1,1}$, and no SUSY in the whole bulk
     space including the throat sector. Nevertheless we can conclude
     that the partition function should vanish in  all the genera. 
     See \cite{KutS3} for the detail.} 

The general solutions $L^{(N-2)}_{l,\bar{l}}$ of (\ref{modinvsu2})
were completely classified by the $A-D-E$ series in ref. \cite{CIZ}. 
In these solutions the values of spin $l/2$ are in 
a one-to-one correspondence
with the exponents of $A-D-E$ Lie algebra, and hence to each of the 
relevant deformations of the singularity $F=0$. 
In this way we obtain the modular invariants classified by 
the $A-D-E$ series corresponding to the singularity type of $X_n$
\cite{OV,DS}.

The appearance of the affine $SU(2)$ character in the expression \eqn{Fl-6-2}
is quite expected. One may relate the background of degenerate K3 surface
to a collection of NS5-branes by means of T-duality \cite{OV,GHM}, 
and it is well-known \cite{CHS} that the world-sheet 
conformal field theory in the NS5 background
contains the $SU(2)$ WZW model.

\subsection{$d=2$ Case}

In the case of $d=2$ the GSO conditions are given as follows;
\begin{itemize}
 \item $NS$-sector
\begin{equation}
F+F_{M_N}+\frac{m}{N}-p Q \in 2\bz +1,
\label{GSO-2-1}
\end{equation}
 \item $R$-sector
\begin{equation}
F+F_{M_N}+\frac{m}{N}-p Q \in 2\bz + 1.
\label{GSO-2-2}
\end{equation}
\end{itemize}
We again determine the spectrum of the momenta $p$ by imposing 
these conditions, 
and the trace for the left-movers is calculated as follows, 
\begin{eqnarray}
\sum_{l=0}^{N-2} G_l&=&\sum_{l=0}^{N-2} \, 
\frac{1}{2}\sum_{m\in\bsz_{2N}}\left\{
\th_3 \ch{(NS)}{lm}(\Th{\frac{m}{N+1}}{\frac{N}{N+1}}
+\Th{\frac{m+N}{N+1}}{\frac{N}{N+1}})
-\th_4 \tch{(NS)}{lm}
(\Th{\frac{m}{N+1}}{\frac{N}{N+1}}
-\Th{\frac{m+N}{N+1}}{\frac{N}{N+1}}) \right.  \nonumber\\
& & \hspace{1in} \left.
-\th_2 \ch{(R)}{lm}
(\Th{\frac{m}{N+1}}{\frac{N}{N+1}}
+\Th{\frac{m+N}{N+1}}{\frac{N}{N+1}}) 
\right\}  .
\end{eqnarray}
We again have $G_l=G_{N-2-l}$ and likewise introduce 
\begin{eqnarray}
  F_l&\df & \frac{1}{2}\sum_{m\in\bsz_{2N}}
\left\{\Th{\frac{m}{N+1}}{\frac{N}{N+1}}
(\th_3 \ch{(NS)}{lm} -\th_4 \tch{(NS)}{lm})
-\Th{\frac{m+N}{N+1}}{\frac{N}{N+1}}\th_2 \ch{(R)}{lm}   \right\},
\label{Fl-2-1} \\
G_l&=& F_l+F_{N-2-l},
\end{eqnarray}
to avoid the double counting of states.
However, it is easy to see that 
$F_l$ here does not have a good modular transformation property.
This is because the theta functions appearing in \eqn{Fl-2-1}
have fractional levels, and they do not close among themselves
under the modular transformation.

In order to avoid this difficulty we further decompose 
$F_l$ into a set of functions $F_{l,r}\hskip1mm (r=0,1,\cdots,2N+1)$
with the help of the formula \eqn{formula} as
\begin{eqnarray}
F_l&=&\sum_{\stackrel{r\in\bsz_{2(N+1)}}{l+r\equiv 0 (\msc{mod }2)}}F_{l,r} ,
\label{Flr-even}\\
F_{l,r}&\df&\frac{1}{2}\sum_{m\in\bsz_{2N}}\Th{(N+1)m+Nr}{N(N+1)}
\left\{\th_3\ch{(NS)}{lm}-(-1)^{l+r}\th_4\tch{(NS)}{lm}-\th_2\ch{(R)}{lm}
\right\}. 
\label{Flr-2}
\end{eqnarray}
$F_{l,r}$ has the symmetry
\begin{equation}
F_{l,r}=F_{l,r+2(N+1)}=F_{N-2-l,r+(N+1)},
\label{symmetry-2}
\end{equation} 
and thus we have
\begin{equation}
F_{N-2-l}=\sum_{\stackrel{r\in\bsz_{2(N+1)}}{l+r\equiv 1(\msc{mod }2)}}F_{l,r}.
\label{Flr-odd}
\end{equation}

It turns out that the functions 
$F_{l,r}(\tau)$ have the good modular properties as;
\begin{eqnarray}
F_{l,r}(\tau+1)&=&
e^{2\pi i\left\{\frac{l(l+2)}{4N}-\frac{N-2}{8N}-\frac{r^2}{4(N+1)}
             +\frac{3+(-1)^{l+r}}{8}  \right\}} \, F_{l,r}(\tau), 
\label{modular Flr-2a}\\
F_{l,r}(-\frac{1}{\tau})&=&(\sqrt{-i\tau})^2 
\sum_{l'=0}^{N-2}\sum_{r'\in\bsz_{2(N+1)}} \, {\cal S}_{(l,r)\,(l',r')}
\,F_{l',r'}(\tau) ,\label{modular Flr-2b} \\
{\cal S}_{(l,r)\,(l',r')} &\df&\frac{(-1)^{l+r}+(-1)^{l'+r'}}{2}
 S^{(N-2)}_{ll'}\frac{1}{\sqrt{N+1}}e^{2\pi i \frac{rr'}{2(N+1)}}. 
\label{modular Flr-2}
\end{eqnarray}
Therefore, it seems reasonable to regard the functions 
$F_{l,r}(\tau)$ as the basic conformal blocks of 
our partition function.
Note that due to \eqn{symmetry-2} the two sets
$\{ F_{l,r} ;~ l+r\equiv 0~ (\mod ~2) \}$ and 
$\{F_{l,r} ;~ l+r\equiv 1~ (\mod ~2)\}$ are not independent and 
we should choose one of these as the building block of the theory.
Let us take the set with $l+r\equiv 0$ $(\mod ~2)$. 
This restriction is consistent since
\eqn{modular Flr-2} implies that the modular transformations 
act separately for each set.

Thanks to the transformation laws \eqn{modular Flr-2a}, \eqn{modular Flr-2b} 
we can now
construct the modular invariant partition function in the following form,
\begin{equation}
\begin{array}{lll}
 Z_{GSO}(\tau,\bar{\tau})&=&
\dsp \frac{1}{|\eta(\tau)|^4}\sum_{l,\bar{l}=0}^{N-2}
 \sum_{r,\bar{r}\in\bsz_{2(N+1)}} \,
 N_{(l,r),(\bar{l},\bar{r})}F_{l,r}(\tau)F_{\bar{l},\bar{r}}(\bar{\tau}) ,
\\
N_{(l,r),(\bar{l},\bar{r})} &=& \dsp  L_{l,\bar{l}}^{(N-2)}
                     M_{r,\bar{r}}^{(N+1)}.
\end{array}
\label{ZGSO-2}
\end{equation}
Here $L_{l,\bar{l}}^{(N-2)}$ again denotes one of 
the $A-D-E$ modular invariants  
of $\widehat{SU}(2)_{N-2}$, and $M_{r,\bar{r}}^{(k)}$ is the modular invariant
of the "theta system" which satisfies the following conditions,
\begin{equation}
\begin{array}{l}
\dsp  M^{(k)}_{r,\bar{r}}=0 ,~~~\mbox{unless } 
\frac{r^2}{4k}-\frac{\bar{r}^2}{4k} \in \bz ,~~~
  M^{(k)}_{r+k,\,\bar{r}+k}=M^{(k)}_{r,\bar{r}} \\
 \dsp \sum_{r,\bar{r}}\, M^{(k)}_{r,\bar{r}}R^{(k)\dag}_{r,r'}
R^{(k)\dag}_{\bar{r},\bar{r}'} = M^{(k)}_{r',\bar{r}'} ,~~~
\dsp R^{(k)}_{r,r'} \df \frac{1}{\sqrt{2k}}e^{-2\pi i\frac{rr'}{2k}}.
 \end{array}
\end{equation}
The simplest solution for $M^{(k)}_{r,\bar{r}}$ is, of course, given by
$M^{(k)}_{r,\bar{r}}=\delta_{r,\bar{r}}$ (or $M^{(N-2)}_{r\bar{r}}
=\delta_{r,-\bar{r}}$),
and the most general solution is given by \cite{GQ} 
\begin{equation}
M^{(k)}_{r,\bar{r}}=\frac{1}{2} \sum_{x\in \bsz_{2\beta},\,y\in\bsz_{2\al}}
\delta_{r,\al x +\beta y}\delta_{\bar{r},\al x-\beta y},
\label{M-GQ}
\end{equation} 
where $\al$, $\beta$ are general integers such that $\al \beta=k$.

A few comments are in order:
\begin{enumerate}
\item Our solution (\ref{ZGSO-2}) 
for the simplest case $N=2$ (the minimal model $M_N$ 
becomes trivial) 
coincides with the one presented in ref.\cite{KutS3}. 
\item It is possible to 
derive the following relations for the functions $F_{l,r}$
by making use of the product formula of theta functions \eqn{product};
\begin{eqnarray}
&&\hskip-25mm \sum_{\stackrel{r\in\bsz_{2(N+1)}}{l+r\equiv 0~(\msc{mod } 2)}}
\Th{r}{N+1}(\tau,0)F_{l,r}(\tau,z) =  \frac{1}{2}\chi^{(N-2)}_l(\tau,0)
\nonumber\\
&& \hspace{2cm} \times
\left\{(\th_3^2-\th_4^2)(\tau, z)\Th{0}{1}(\tau,2z) 
-(\th_2^2+\th_1^2)(\tau,z)\Th{1}{1}(\tau,2z)\right\}, \label{identity-2a}\\
&&\hskip-25mm 
\sum_{\stackrel{r\in\bsz_{2(N+1)}}{l+r\equiv 1~(\msc{mod } 2)}}
\Th{r}{N+1}(\tau,0)F_{l,r}(\tau,z) =  \frac{1}{2}\chi^{(N-2)}_l(\tau,0)
\nonumber\\
&& \hspace{2cm} \times
\left\{(\th_3^2+\th_4^2)(\tau, z)\Th{1}{1}(\tau,2z)  
-(\th_2^2-\th_1^2)(\tau,z)\Th{0}{1}(\tau,2z)\right\}, \label{identity-2b}\\
&&F_{l,r}(\tau,z)\df\frac{1}{2}\sum_{m\in\bsz_{2N}}\Th{(N+1)m+Nr}{N(N+1)}
 (\tau, -\frac{2z}{N}) \nonumber \\
&&~~~~~ \times
\left(\th_3\ch{(NS)}{lm}-(-1)^{l+r}\th_4\tch{(NS)}{lm}-\th_2\ch{(R)}{lm}
-i(-1)^{l+r}\th_1\tch{(R)}{lm}\right)(\tau,z).
\label{identity-2}
\end{eqnarray}
(Note $F_{l,r}(\tau)\equiv F_{l,r}(\tau,z=0)$). 
It is known \cite{BG} that the combination of 
theta functions in the right-hand-side of 
(\ref{identity-2a}),(\ref{identity-2b}) vanishes identically
\begin{eqnarray}
&& (\th_3^2-\th_4^2)(\tau, z)\Th{0}{1}(\tau,2z)
   -(\th_2^2+\th_1^2)(\tau, z)\Th{1}{1}(\tau,2z)=0 , \\
&& (\th_3^2+\th_4^2)(\tau, z)\Th{1}{1}(\tau,2z)
   -(\th_2^2-\th_1^2)(\tau, z)\Th{0}{1}(\tau,2z)=0 . 
\end{eqnarray} 

Thus the sum of functions $F_{l,r}$ 
(\ref{identity-2a}), (\ref{identity-2b}) in fact vanishes identically.
Then these equations imply that the functions $F_{l,r}$ themselves
should vanish separately for each $|r|$ 
since the level-($N+1$) theta functions 
$\Theta_{r,N+1}(\tau,0)$ are functionally independent
for different $|r|$.
We have explicitly verified by Maple that 
$F_{l,r}(\tau,z)+F_{l,-r}(\tau,z)$ in fact vanishes 
in lower orders in $q\equiv e^{2\pi i \tau}$, 
$y \equiv e^{2\pi i z}$ and $y^{-1}$, for every $l$, $r$ and 
$N=2,3,4$. 

We conjecture that the identity
\begin{equation} 
F_{l,r}(\tau,z)+F_{l,-r}(\tau,z) \equiv 0
\end{equation}
holds for arbitrary $l$, $r$, $N$. If this is the case, we have
$F_{l,r}(\tau)\equiv F_{l,r}(\tau,0)\equiv F_{l,-r}(\tau,0)=0$ and the
partition function vanishes $Z_{GSO}=0$: 
this is consistent with the presence of the space-time supersymmetry.

\item The modular properties of 
$F_{l,r}(\tau)$ \eqn{modular Flr-2a}, \eqn{modular Flr-2b}
can be immediately read off from \eqn{identity-2a}, \eqn{identity-2b}:
we find that the index $l$ of $F_{l,r}$ transforms like the spin of the
representation of 
affine $SU(2)$ and the index $r$ transforms like a label of the
$U(1)$ theta function. Modular properties of $F_{l,r}(\tau)$ is in fact
similar to those of parafermionic theory \cite{GQ}.

\item $F_{l,r}(\tau,z)$ is transformed 
under the spectral flow $\dsp z\mapsto  z+\frac{\al}{2}\tau$ 
$(\al \in \bz)$ \cite{LVW,EOTY} as follows;
\begin{equation}
F_{l,r}(\tau,z+\frac{\al}{2}\tau)=(-1)^{\al}q^{-\frac{\al^2}{2}}y^{-2\al}
F_{l,r}(\tau,z).
\end{equation}
This means that $F_{l,r}(\tau,z)$ is a "flow-invariant orbit"
in the sense of \cite{EOTY}. This fact justifies regarding $F_{l,r}$
as the building block of the partition function.
\end{enumerate}

\subsection{$d=4$ Case}

The GSO conditions are given as follows;
\begin{itemize}
 \item $NS$-sector
\begin{equation}
F+F_{M_N}+\frac{m}{N}-p Q \in 2\bz +1,
\label{GSO-4-1}
\end{equation}
 \item $R$-sector
\begin{equation}
F+F_{M_N}+\frac{m}{N}-p Q \in 2\bz + \frac{1}{2}.
\label{GSO-4-2}
\end{equation}
\end{itemize}

In this case the  trace over  the left-moving  Hilbert space is given by, 
\begin{eqnarray}
 \sum_{l=0}^{N-2} G_l&=&\sum_{l=0}^{N-2} 
\frac{1}{2}\sum_{m\in\bsz_{2N}}\left\{
\th_3^2\ch{(NS)}{lm}(\Th{\frac{2m}{N+2}}{\frac{2N}{N+2}}
+\Th{\frac{2(m+N)}{N+2}}{\frac{2N}{N+2}})
-\th_4^2\tch{(NS)}{lm}
(\Th{\frac{2m}{N+2}}{\frac{2N}{N+2}}
-\Th{\frac{2(m+N)}{N+2}}{\frac{2N}{N+2}}) \right.  \nonumber\\
& & \hspace{1in} \left.
-\th_2^2\ch{(R)}{lm}
(\Th{\frac{2m+N}{N+2}}{\frac{2N}{N+2}}
+\Th{\frac{2m-N}{N+2}}{\frac{2N}{N+2}})
\right\}.
\end{eqnarray}
Again $G_l$ consists of theta functions
with fractional levels and we have $G_l=G_{N-2-l}$.
We introduce $\dsp F_l= \frac{1}{2}G_l$
and expand $F_l$ into a set of functions $F_{l,r}\hskip1mm
(r=0,1,\cdots,2N+3)$,
\begin{eqnarray}
F_{l,r}&\df&\frac{1}{4}\sum_{m\in\bsz_{4N}}\Th{(N+2)m+Nr}{2N(N+2)}
\left\{\th_3^2\ch{(NS)}{lm}-(-1)^{\frac{r+m}{2}}\th_4^2\tch{(NS)}{lm}
-\th_2^2\ch{(R)}{lm}  \right\}  \\ \nonumber
&&\hspace{4cm} (l+r\equiv 0~(\mod~2)), \nonumber\\
F_{l,r}&\df&0, \hspace{3cm} (l+r\equiv 1~(\mod~2)), \nonumber\\
F_l&=& \sum_{r\in\bsz_{2(N+2)}} F_{l,r}.
\label{Flr-4} 
\end{eqnarray}
It is easy to see that
\begin{equation}
F_{l,r}=F_{l,\, r+2(N+2)}=F_{N-2-l,\, r+(N+2)}.
\label{symmetry-4}
\end{equation}
Note that \eqn{symmetry-4} is consistent with 
the definition $F_{l,r}\equiv 0 $ $(l+r\equiv 1 ~(\mod~ 2))$,
since $l+r\equiv l+r+2(N+2) \equiv (N-2-l)+(r+N+2) ~(\mod~ 2)$.

$F_{l,r}$ possess the following modular transformation properties;
\begin{eqnarray}
F_{l,r}(\tau+1)&=&
e^{2\pi i\left\{\frac{l(l+2)}{4N}-\frac{N-2}{8N}-\frac{r^2}{4(N+2)}
             +\frac{1}{2}  \right\}} \, F_{l,r}(\tau) , \\
F_{l,r}(-\frac{1}{\tau})&=&(\sqrt{-i\tau})^3 
\sum_{l'=0}^{N-2}\sum_{r'\in\bsz_{2(N+2)}} \, {\cal S}_{(l,r)\,(l',r')}
\,F_{l',r'}(\tau) , \\
{\cal S}_{(l,r)\,(l',r')} &\df&
 S^{(N-2)}_{ll'}\frac{1}{\sqrt{2(N+2)}}e^{2\pi i \frac{rr'}{2(N+2)}} ,
\label{modular Flr-4}
\end{eqnarray}
It is now easy to construct modular invariant partition functions
\begin{equation}
\begin{array}{lll}
 Z_{GSO}(\tau,\bar{\tau}) 
&=&\dsp \frac{1}{|\eta(\tau)|^6}\sum_{l,\bar{l}=0}^{N-2}
 \sum_{r,\bar{r}\in\bsz_{2(N+2)}}\, 
N_{(l,r),(\bar{l},\bar{r})}F_{l,r}(\tau)F_{\bar{l},\bar{r}}(\bar{\tau}) ,
\\
N_{(l,r),(\bar{l},\bar{r})} &=& \dsp \frac{1}{2}\left(L_{l,\bar{l}}^{(N-2)}
                     M_{r,\bar{r}}^{(N+2)}+ L_{N-2-l,\,\bar{l}}^{(N-2)}
                     M_{r+N+2, \, \bar{r}}^{(N+2)} \right) ,
\end{array}
\label{ZGSO-4}
\end{equation}
where $L^{(k)}_{l,\bar{l}}$ and $M^{(k)}_{r,\bar{r}}$ 
are defined as before.

The solution for the simplest case $N=2$ 
(the case of conifold singularity in $CY_3$ \cite{GV}) 
was first obtained by S. Mizoguchi from a somewhat 
different approach \cite{Mizoguchi}.

As in the two-dimensional case, we can construct the following combination
of the $F_{l,r}$ functions
\begin{eqnarray}
&& \sum_{r\in\bsz_{2(N+2)}}\Th{r}{N+2}(\tau,0)
  F_{l,r}(\tau,z)= \frac{1}{4}
\chi^{(N-2)}_l (\tau,0)\left(\th_3^4(\tau,z)-\th_4^4(\tau,z) 
 -\th_2^4(\tau,z)+\th_1^4(\tau,z)\right), \label{identity-4a}\\
&&  F_{l,r}(\tau,z)\df \frac{1}{4}\sum_{m\in\bsz_{4N}}
 \Th{(N+2)m+Nr}{2N(N+2)}(\tau,-\frac{z}{N}) \nonumber \\
&&  ~~~\times
 \left(\th_3^2\ch{(NS)}{lm}-(-1)^{\frac{r+m}{2}}\th_4^2\tch{(NS)}{lm}
 -\th_2^2\ch{(R)}{lm}  +i(-1)^{\frac{r+m}{2}}\th_1^2\tch{(R)}{lm} \right)
(\tau,z) , ~~~(l+r\equiv 0~(\mbox{mod}~2)) \nonumber \\
&& F_{l,r}(\tau,z)\df 0 , \hspace{1cm}(l+r\equiv 1~(\mbox{mod}~2)).
\label{identity-4}
\end{eqnarray}
We note that the right-hand-side of \eqn{identity-4a} vanishes due to 
Jacobi's identity. 
Then as in the case of two-dimensional theories, we expect that functions
$F_{l,r}$ should vanish separately for each $|r|$, $F_{l,r}(\tau,z)+
F_{l,-r}(\tau,z)\equiv 0$. We have explicitly checked this for 
lower  orders of $q$, $y$, $y^{-1}$ by Maple and found that in fact a stronger 
relation
\begin{equation}
F_{l,r}(\tau, z)\equiv 0, 
\label{vanishF4}
\end{equation}
holds. We conjecture that (\ref{vanishF4}) holds for all $l,r,N$.
In this case all the modular invariant theories again have vanishing
partition functions 
and are consistent with the presence of space-time supersymmetry.

We may again read off the modular 
transformation rule \eqn{modular Flr-4} from
the identity \eqn{identity-4a}: $F_{l,r}$ transforms like
an affine $\widehat{SU}(2)$ character in its index $l$ and like $U(1)$ 
theta function in its label $r$.

We may show
\begin{equation}
F_{l,r}(\tau,z+\frac{\al}{2}\tau)=(-1)^{\al}q^{-\frac{\al^2}{2}}y^{-2\al}
F_{l,r}(\tau,z),
\end{equation}
which implies that the functions $F_{l,r}(\tau,z)$ are 
the flow-invariant orbits for each $l,r$.

\bigskip

\section{Conclusions}

\hspace*{4.5mm}

In this paper we have constructed the toroidal partition
functions of the non-critical superstring theory on
$\br^{d-1,1}\times\left(\br_{\phi}\times S^1_Y\right)\times
M_N$, which is to provide the dual description of the singular
Calabi-Yau compactification in the decoupling limit.
We have found that there exists a natural $A-D-E$ classification
of modular invariants associated to the type of Calabi-Yau singularities
in all cases of $d=6,4,2$. In cases 
$d=4,2$, we found that the conformal blocks 
composing the partition function behave 
like primary fields of the parafermionic theory.
It will be very interesting if we could identify our conformal blocks with 
suitable scaling operators in respective field theories and elucidate
their dynamical properties.

As we have discussed at the beginning of Section 3, the presence of the
background charge in the Liouville sector creates a gap $h\ge Q^2/8$ 
in the CFT spectrum. In particular the graviton (which corresponds to $h=0$)
does not appear in the modular invariant partition function.
Thus the system in fact describes some non-gravitational theory 
and the theory is interpreted as being 
at the decoupling limit of type II superstring.
It is quite reassuring to us that one can construct modular invariant 
amplitudes for string propagation even in such a "singular" situation where
some of the conventional world-sheet technology may break down and 
non-perturbative effects play an important role. 

Landau-Ginzburg theory has the disturbing feature of the 
appearance of a negative power piece in the superpotential when applied to
describe singular (non-compact) Calabi-Yau manifolds. It is not clear how to 
treat the negative power operator within the framework of the standard
${\cal N}=2$ SCFT. It now appears, however, the negative power term 
may be handled properly by means of the Liouville degrees of freedom
with an appropriate background charge.
The appearance of the gap and the continuous spectrum above the gap in string
propagation in singular Calabi-Yau manifold are reproduced 
exactly by the dynamics of the Liouville field. It will be extremely
useful if we have a better understanding on the relationship between 
the singular geometry and the dynamics of 
Liouville field.

It will be interesting to consider more general class of 
${\cal N}=2$ models instead of ${\cal N}=2$ minimal model
(Landau-Ginzburg orbifolds \cite{LGO} or Gepner models \cite{Gepner}, etc.). 
Quite recently, in ref. \cite{Pelc}, 
Landau-Ginzburg orbifolds are discussed, 
relating it to the ${\cal N}=2$ $SCFT_4$ 
{\em with matter fields\/} \cite{AD}. 
It may also be interesting to study
non-rational Calabi-Yau singularities (collapse of del Pezzo surfaces etc.). 
These problems may be regarded as natural generalizations
of the Gepner model, namely, the (orbifoldized) tensor product of  
minimal models ("compact models") with the Liouville theory
("non-compact" models). Construction of
modular invariants for such models
will provide important consistency checks of their dynamics.

\bigskip

\bigskip

\section*{Acknowledgement}

We would like to thank especially Dr. S.Mizoguchi whose talk at Univ. of 
Tokyo stimulated the present investigation. 
Y.S. would also thank Drs. K.Ito and A.Kato for useful
comments, and Prof. I.Bars and his theory group
for kind hospitality at USC. Part of this work was done 
while Y.S. was attending the workshop 
"Strings, Branes and M-theory" at 
CIT-USC Center for Theoretical Physics.  

This work is supported in part by 
Grant-in-Aid for Scientific Research on Priority Area 
$\sharp 707$ "Supersymmetry and Unified Theory
of Elementary Particles" from Japan Ministry of Education.

\newpage
\appendix
\section{Notations and Conventions}

In this appendix we summarize our conventions and present some formulas 
used in the manuscript. 

\begin{description}
\item[1. theta functions] 

\bigskip

We set $q:= e^{2\pi i \tau}$, $y:=e^{2\pi i z}$ and introduce 
various theta functions;
 \begin{equation}
 \begin{array}{l}
 \dsp \th_1(\tau,z) =i\sum_{n=-\infty}^{\infty}(-1)^n q^{(n-1/2)^2/2} y^{n-1/2}
  \equiv 2 \exp(\frac{\pi i \tau}{4})\sin(\pi z)\prod_{m=1}^{\infty}
    (1-q^m)(1-yq^m)(1-y^{-1}q^m), \\
 \dsp \th_2(\tau,z)=\sum_{n=-\infty}^{\infty} q^{(n-1/2)^2/2} y^{n-1/2}
  \equiv 2 \exp(\frac{\pi i \tau}{4})\cos(\pi z)\prod_{m=1}^{\infty}
    (1-q^m)(1+yq^m)(1+y^{-1}q^m), \\
 \dsp \th_3(\tau,z)=\sum_{n=-\infty}^{\infty} q^{n^2/2} y^{n}
  \equiv \prod_{m=1}^{\infty}
    (1-q^m)(1+yq^{m-1/2})(1+y^{-1}q^{m-1/2}), \\
 \dsp \th_4(\tau,z)=\sum_{n=-\infty}^{\infty}(-1)^n q^{n^2/2} y^{n}
  \equiv \prod_{m=1}^{\infty}
    (1-q^m)(1-yq^{m-1/2})(1-y^{-1}q^{m-1/2}) .
 \end{array}
 \end{equation}
 \begin{equation}
 \Th{m}{k}(\tau,z)=\sum_{n=-\infty}^{\infty}
 q^{k(n+\frac{m}{2k})^2}y^{k(n+\frac{m}{2k})} .
 \end{equation}
We use the abbreviations; $\th_i \equiv \th_i(\tau, 0)$
($\th_1\equiv 0$), $\Th{m}{k}(\tau) \equiv \Th{m}{k}(\tau,0)$.
We also define
\begin{equation}
\eta(\tau)=q^{1/24}\prod_{n=1}^{\infty}(1-q^n)
\end{equation}

The product formula of theta function is written as \cite{Kac,KYY};
\begin{equation}
\Th{m}{k}(\tau,z)\Th{m'}{k'}(\tau,z')
=\sum_{r\in\bsz_{k+k'}}\Th{mk'-m'k+2kk'r}{kk'(k+k')}(\tau,u)
\Th{m+m'+2kr}{k+k'}(\tau,v),
\label{product}
\end{equation} 
where we set 
$\dsp u= \frac{z-z'}{k+k'}$, $\dsp v=\frac{kz+k'z'}{k+k'}$.

The following identity is often used ($p$ is an integer);
\begin{equation}
\Th{m/p}{k/p}(\tau,z)=\Th{m}{k}(\tau/p,z/p)=
\sum_{r\in\bsz_p} \,\Th{m+2kr}{pk}(\tau,z/p).
\label{formula}
\end{equation}

\item[2. characters of ${\cal N}=2$ minimal model]

\bigskip

Let $\chi^{(k)}_l(\tau, z)$ be the 
character of $\widehat{SU}(2)_k$ with the spin $l/2$ ($0\leq l \leq k$)
representation;
\begin{equation}
\chi^{(k)}_l(\tau, z) =\frac{\Th{l+1}{k+2}-\Th{-l-1}{k+2}}
                        {\Th{1}{2}-\Th{-1}{2}}(\tau, z) ~  .
\end{equation}
String function $c^l_m(\tau)$ is defined by
\begin{equation}
\chi^{(k)}_l(\tau, z)= \sum_{m\in \bsz_{2k}}c^l_m(\tau)\Th{m}{k}(\tau,z).
\end{equation}
We introduce
\begin{equation}
\chi_m^{l,s}(\tau,z)=\sum_{r\in \bsz_k}c^l_{m-s+4r}(\tau)
\Th{2m+(k+2)(-s+4r)}{2k(k+2)}(\tau,z/(k+2)).
\end{equation}
String function $c^l_m$ has the following properties;
\begin{equation}
c^l_m=c^l_{-m}=c^l_{m+2k}=c^{k-l}_{m+k}, ~~~c^l_m=0 ~\mbox{unless }
  l+m\equiv 0 ~(\mod ~ 2).
\end{equation}
Likewise, $\chi^{l,\,s}_m$ has the properties;
\begin{equation}
\chi^{l,\, s}_m=\chi^{l,s}_{m+2(k+2)}=\chi^{l\,s+4}_m=
\chi^{k-l,\, s+2}_{m+(k+2)}, ~~~\chi^{l,\, s}_m=0 ~\mbox{unless }
  l+m+s \equiv 0 ~(\mod ~ 2),
\end{equation}
and thus $m$, $s$ run over the range $m\in \bz_{2(k+2)}$, $s\in\bz_4$.

The characters of ${\cal N}=2$ minimal model with 
$\dsp \hat{c}=\frac{k}{k+2}$ are defined \cite{RY,Gepner}  by 
\begin{equation}
\begin{array}{l}
 \dsp \ch{(NS)}{l,m}(\tau,z) 
 \equiv \tr_{{\cal H}^{NS}_{l,m}} q^{L_0-\hat{c}/8}y^{J_0}
 = \chi^{l,0}_m + \chi^{l,2}_m \\
\dsp \tch{(NS)}{l,m}(\tau,z)
  \equiv \tr_{{\cal H}^{NS}_{l,m}} (-1)^F q^{L_0-\hat{c}/8}y^{J_0}
 = \chi^{l,0}_m - \chi^{l,2}_m \\
\dsp  \ch{(R)}{l,m} (\tau,z)
\equiv \tr_{{\cal H}^{R}_{l,m}} q^{L_0-\hat{c}/8}y^{J_0}
 = \chi^{l,1}_m + \chi^{l,3}_m \\
\dsp \tch{(R)}{l,m} (\tau,z)
\equiv \tr_{{\cal H}^{R}_{l,m}}(-1)^F q^{L_0-\hat{c}/8}y^{J_0}
 = \chi^{l,1}_m - \chi^{l,3}_m .
\end{array}
\end{equation}

It is easy to prove  the following branching relation
by means of the product formula of theta functions 
\eqn{product} \cite{Gepner,KYY};
\begin{equation}
\chi_l^{(k)}(\tau,w)\Th{s}{2}(\tau,w-z)
=\sum_{m\in \bsz_{2(k+2)}} \chi_m^{l,s}(\tau,z)\Th{m}{k+2}(\tau,w-2z/(k+2)).
\label{branching}
\end{equation}
This relation \eqn{branching} represents the
minimal model as the Kazama-Suzuki coset for
$\dsp SU(2)_k/U(1)$.

\end{description}

\section{Equivalence between the Kazama-Suzuki Model for $SL(2;\br)/U(1)$
and the ${\cal N}=2$ Liouville Theory}

\hspace*{4.5mm}

In this appendix we discuss  the equivalence between
the ${\cal N}=2$ coset SCFT for $SL(2;\br)/U(1)$ and the ${\cal N}=2$ Liouville theory 
from the viewpoint of free field realizations.
We start from the following free field realization\footnote
   {We have another familiar free field realization;
    "Wakimoto representation" \cite{Wakimoto} $(\varphi,\,\beta,\,\gamma)$.
     The relation between these free fields and $X,\,\phi,\,u$ here
     is given as follows;
$$
\left\{
\begin{array}{l}
 \dsp \varphi =\phi + \sqrt{\frac{k}{k+2}}(u+iX)  ,\\
 \dsp \beta=-\left(\sqrt{\frac{k+2}{2}}i\partial X
  +\sqrt{\frac{k}{2}}\partial \phi\right)e^{\sqrt{\frac{2}{k+2}}(u+iX)} ,\\
 \dsp \gamma=e^{-\sqrt{\frac{2}{k+2}}(u+iX)} .
\end{array}
\right.
$$  } of
$SL(2;\br)$-current algebra with the level $k+2$,
\begin{equation}
\left\{
\begin{array}{l}
 \dsp J^3=\sqrt{\frac{k+2}{2}}\partial u \\
 \dsp J^{\pm}=-\left(\sqrt{\frac{k+2}{2}}i\partial X
  \mp \sqrt{\frac{k}{2}}\partial \phi\right)
   e^{\mp\sqrt{\frac{2}{k+2}}(u+iX)} ,
\end{array}
\right.
\end{equation}
where $X(z)X(0)\sim -\ln z$, $\phi(z)\phi(0)\sim -\ln z$, 
$u(z)u(0)\sim -\ln z$ are free scalar fields.
The Sugawara stress tensor is given by 
\begin{equation}
T_{SL(2;\bsr)}=-\frac{1}{2}(\partial X)^2
-\frac{1}{2}(\partial \phi)^2 -\frac{1}{\sqrt{2k}}\partial^2\phi
-\frac{1}{2}(\partial u)^2 ,
\end{equation}  
which possesses the central charge $\dsp c=3+\frac{6}{k}$.
($k$ is related to $N$ in the text as $k=N,\hskip1mm d=6; k=2N/(N+2), 
\hskip1mmd=4; k=N/(N+1), \hskip1mm d=2$, respectively).

The Kazama-Suzuki model \cite{KazS} for $SL(2;\br)/U(1)$ is 
given by further tensoring the system with
two $U(1)$-charged fermions;
$\dsp \psi^+(z)\psi^-(0)\sim \frac{1}{z}$, and then by 
gauging the $U(1)$-subgroup. 
Here we adopt the BRST formulation and first 
bosonize the $U(1)$-gauge field as $A(z)\sim i\partial v(z)$,
where $v(z)$ is a real scalar field with 
$\dsp v(z)v(0)\sim -\ln z$. We also bosonize the 
fermions $\psi^{\pm}$ in the standard fashion;
\begin{equation}
\psi^{\pm}(z)=e^{\pm iH(z)},~~~H(z)H(0)\sim -\ln z.
\end{equation}

The total stress tensor for the Kazama-Suzuki model then reduces to
the following form,
\begin{equation}
T = -\frac{1}{2}(\partial X)^2
-\frac{1}{2}(\partial \phi)^2 -\frac{1}{\sqrt{2k}}\partial^2\phi
-\frac{1}{2}(\partial u)^2 -\frac{1}{2}(\partial v)^2
-\frac{1}{2}(\partial H)^2 -\eta\partial \xi, 
\label{T-KS}
\end{equation}
where $(\xi, \eta)$ is  the spin (0,1) ghost system
and the BRST charge is given by
\begin{equation}
Q_{U(1)}=\oint\xi\left(\sqrt{\frac{k+2}{2}}\partial u 
+i\sqrt{\frac{k}{2}}\partial v + i\partial H\right).
\label{BRST-KS}
\end{equation}
This stress tensor \eqn{T-KS} has the correct central charge
$\dsp c=3(1+\frac{2}{k})$ and
the world-sheet ${\cal N}=2$ superconformal symmetry 
is generated by the currents,
\begin{equation}
\left\{
\begin{array}{l}
 \dsp G^{\pm}=\frac{1}{\sqrt{k}}\psi^{\pm}J^{\mp}
=-\frac{1}{\sqrt{k}}
\left(\sqrt{\frac{k+2}{2}}i\partial X\pm \sqrt{\frac{k}{2}}\partial \phi\right)
e^{\pm \sqrt{\frac{2}{k+2}}(u+iX)\pm iH}, \\
\dsp J=\psi^+\psi^-+\frac{2}{k}(J^3+\psi^+\psi^-)=\frac{k+2}{k}i\partial H
 +\frac{\sqrt{2(k+2)}}{k}\partial u .
\end{array}
\right.
\label{SCA1}
\end{equation}

Now, let us try to reduce the above Kazama-Suzuki model to the 
${\cal N}=2$ Liouville theory. For this purpose it is convenient to introduce 
the following field redefinition,
\begin{equation}
\left(
\begin{array}{c}
 v'\\ H'
\end{array}
\right) \df
\left(
\begin{array}{cc}
 \sqrt{\frac{k}{k+2}} & \sqrt{\frac{2}{k+2}} \\
 -\sqrt{\frac{2}{k+2}} & \sqrt{\frac{k}{k+2}} 
\end{array}
\right)
\left(
\begin{array}{c}
 v \\ H
\end{array}
\right) .
\end{equation}
Clearly we have $v'(z)v'(0)\sim -\ln z$,  $H'(z)H'(0)\sim -\ln z$
and $v'(z)H'(0)\sim 0$.
The BRST-charge \eqn{BRST-KS} is rewritten as
\begin{equation}
Q_{U(1)}=\sqrt{\frac{k+2}{2}}\oint \xi(\partial u+i\partial v'),
\end{equation} 
and the stress tensor \eqn{T-KS} is reexpressed as follows,
\begin{equation}
T=-\frac{1}{2}(\partial X)^2 
-\frac{1}{2}(\partial \phi)^2 -\frac{1}{\sqrt{2k}}\partial^2\phi
-\frac{1}{2}(\partial H')^2 +\left\{ Q_{U(1)},~ 
\frac{1}{\sqrt{2(k+2)}}\eta(-\partial u+i\partial v')\right\}.
\end{equation}
We also obtain the following expressions for $G^{\pm}$, $J$, 
\begin{equation}
\left\{
\begin{array}{l}
\dsp  G^{\pm}= -\frac{1}{\sqrt{k}}\left(\sqrt{\frac{k+2}{2}}
 i\partial X\pm \sqrt{\frac{k}{2}}\partial \phi\right) 
 e^{\pm i\left(\sqrt{\frac{k}{k+2}}H'+\sqrt{\frac{2}{k+2}}X \right)
        \pm \sqrt{\frac{2}{k+2}} (u+iv')}, \\
\dsp  J= \sqrt{\frac{k+2}{k}}i\partial H' 
+ \left\{Q_{U(1)},~ \frac{2}{k} \eta \right\} .
\end{array}
\right.
\end{equation}
In order to eliminate $u$, $v'$ in these formulas 
we set $\dsp U\df e^{-\sqrt{\frac{2}{k+2}}\oint(u+iv')J}$ and perform
the similarity transformation,
\begin{equation}
G^{'\pm} \df UG^{\pm}U^{-1} =
-\frac{1}{\sqrt{k}}\left(\sqrt{\frac{k+2}{2}}
 i\partial X\pm \sqrt{\frac{k}{2}}\partial \phi\right) 
 e^{\pm i\left(\sqrt{\frac{k}{k+2}}H'+\sqrt{\frac{2}{k+2}}X \right)}.
\end{equation}
Stress tensor and $U(1)$ current remain invariant
$T'\df UTU^{-1}=T$, $ J'\df UJU^{-1}=J$ 
up to $Q_{U(1)}$-exact terms due to following relations, 
\begin{equation}
\begin{array}{lll}
\dsp \left[\,-\sqrt{\frac{2}{k+2}}\oint (u+iv')J,~J(z)\right] &= &\dsp \left\{
 Q_{U(1)},~ -\frac{2}{k}\eta(z)\right\}, \\
\dsp \left[\, -\sqrt{\frac{2}{k+2}}\oint (u+iv')J,~T(z)\right] &= &0
\end{array}
\end{equation}
It is also obvious that $UQ_{U(1)}U^{-1}=Q_{U(1)}$, and hence 
this similarity transformation is in fact well-defined on the Hilbert
space of Kazama-Suzuki model.
Furthermore it is convenient to rotate again  $X$, $H'$ as,  
\begin{equation}
\left(
\begin{array}{c}
 Y \\ H''
\end{array}
\right) \df
\left(
\begin{array}{cc}
 \sqrt{\frac{k}{k+2}} & -\sqrt{\frac{2}{k+2}} \\
 \sqrt{\frac{2}{k+2}} & \sqrt{\frac{k}{k+2}} 
\end{array}
\right)
\left(
\begin{array}{c}
 X \\ H'
\end{array}
\right) ,
\end{equation}
and set $\Psi^{\pm}\df e^{\pm i H''}$.
Then we finally obtain (up to $Q_{U(1)}$-exact terms)
\begin{equation}
\left\{
\begin{array}{l}
\dsp  T'=-\frac{1}{2}(\partial Y)^2 
-\frac{1}{2}(\partial \phi)^2 -\frac{1}{\sqrt{2k}}\partial^2\phi
-\frac{1}{2}(\Psi^+\partial \Psi^- -\partial \Psi^+ \Psi^-) \\
\dsp  G^{'\pm}=-\frac{1}{\sqrt{2}}\Psi^{\pm}(i\partial Y \pm \partial \phi )
\mp \frac{1}{\sqrt{k}}\partial \Psi^{\pm} \\
\dsp J'= \Psi^+\Psi^- -\sqrt{\frac{2}{k}}i\partial Y.
\end{array}
\right.
\label{SCA2}
\end{equation}
This is no other than the superconformal currents of the ${\cal N}=2$
Liouville theory $\br_{\phi}\times S^1$ with the background charge 
$\dsp Q= \sqrt{\frac{2}{k}}$.
Notice also that the Liouville potential  
$\dsp \delta S_{\pm}=\mu \int d^2z\, \Psi^{\pm}\bar{\Psi}^{\pm}
e^{-\frac{1}{Q}(\phi\mp i Y)}$
is actually a screening operator (
it commutes with all of the superconformal currents \eqn{SCA2})
and moreover $U(\delta S_{\pm}) U^{-1}=\delta S_{\pm}$ holds.
Thus we do not have to make modification in the above derivation
of equivalence, even in the presence of such an interaction term.    


\end{document}